\documentstyle[12pt]{article}
\def\hybrid{\topmargin 0pt      \oddsidemargin 0pt
        \headheight 0pt \headsep 0pt
        \textheight 9in         
        \textwidth 6.25in }      
\catcode`\@=11
\def\marginnote#1{}
\newcount\hour
\newcount\minute
\newtoks\amorpm
\hour=\time\divide\hour by60
\minute=\time{\multiply\hour by60 \global\advance\minute by-\hour}
\edef\standardtime{{\ifnum\hour<12 \global\amorpm={am}%
        \else\global\amorpm={pm}\advance\hour by-12 \fi
        \ifnum\hour=0 \hour=12 \fi
        \number\hour:\ifnum\minute<10 0\fi\number\minute\the\amorpm}}
\edef\militarytime{\number\hour:\ifnum\minute<10 0\fi\number\minute}
\def\draftlabel#1{{\@bsphack\if@filesw {\let\thepage\relax
   \xdef\@gtempa{\write\@auxout{\string
      \newlabel{#1}{{\@currentlabel}{\thepage}}}}}\@gtempa
   \if@nobreak \ifvmode\nobreak\fi\fi\fi\@esphack}
        \gdef\@eqnlabel{#1}}
\def\@eqnlabel{}
\def\@vacuum{}
\def\draftmarginnote#1{\marginpar{\raggedright\scriptsize\tt#1}}
\def\draft{\oddsidemargin -.5truein
        \def\@oddfoot{\sl preliminary draft \hfil
        \rm\thepage\hfil\sl\today\quad\militarytime}
        \let\@evenfoot\@oddfoot \overfullrule 3pt
        \let\label=\draftlabel
        \let\marginnote=\draftmarginnote
   \def\@eqnnum{(\theequation)\rlap{\kern\marginparsep\tt\@eqnlabel}%
\global\let\@eqnlabel\@vacuum}  }
\def\numberbysection{\@addtoreset{equation}{section}
        \def\theequation{\thesection.\arabic{equation}}}
\def\underline#1{\relax\ifmmode\@@underline#1\else
        $\@@underline{\hbox{#1}}$\relax\fi}
\def\titlepage{\@restonecolfalse\if@twocolumn\@restonecoltrue\onecolumn
     \else \newpage \fi \thispagestyle{empty}\c@page\z@
        \def\thefootnote{\fnsymbol{footnote}} }
\def\endtitlepage{\if@restonecol\twocolumn \else  \fi
        \def\thefootnote{\arabic{footnote}}
        \setcounter{footnote}{0}}  
\catcode`@=12
\relax
\def\ie{\hbox{\it i.e. }}        \def\etc{\hbox{\it etc.}}
\def\beq{\begin{equation}}
\def\eeq{\end{equation}}
\relax
\hybrid
\begin{document}
\begin{titlepage}
\begin{center}
April~1997 \hfill    PAR--LPTHE 97-13\\
\hfill cond-mat/9704221\\[.4in]
{\large\bf Weak Randomness for large $q$-State Potts models in 
Two Dimensions}\\[.3in] 
	{\bf M. Picco}\\
	{\it LPTHE\/}\footnote{Laboratoire associ\'e No. 280 au CNRS}\\
       \it  Universit\'e Pierre et Marie Curie, PARIS VI\\
       \it Universit\'e Denis Diderot, PARIS VII\\
	Boite 126, Tour 16, 1$^{\it er}$ \'etage \\
	4 place Jussieu\\
	F-75252 Paris CEDEX 05, FRANCE\\
\end{center}
\vskip .1in
\centerline{\bf ABSTRACT}
\begin{quotation}
We have studied the effect of weak randomness on $q$-state Potts models for
$q> 4$ by measuring the central charges of these models using transfer
matrix methods. We obtain a set of new values for the central charges and
then show that some of these values are related to one another by a
factorization law. 
\end{quotation}

\end{titlepage}
\newpage

The effect of weak randomness on second order phase transitions has been
the subject of many recent works. These studies were either analytical
\cite{dd1,lud1,dpp} or numerical \cite{adsw,mp,dw}. Recently, models
with first order phase transitions have also been investigated. According
to general arguments, it is expected that a model with a first order phase
transition will behave like a second order phase transition in presence of
weak disorder \cite{iw,hb}. This has been put on a more rigorous level in
\cite{aw} and checked numerically in \cite{cfl} where the $8$-state Potts
model was studied in presence of disorder. In this work, it was also found
that the critical exponents of the second order phase transitions are the
ones of the Ising model. In a recent study, Cardy considered the case of
$N$ coupled Ising models with weak disorder \cite{cardy}. By considering on
an equal footing the coupling between the $N$ Ising models and the
disorder, Cardy was able to show that this model flows to $N$ decoupled
Ising models. This study was extended to the case of $N$ coupled $3$-state
Potts models with disorder, with similar conclusions \cite{pujol}.

Based on these results, it was suggested by Cardy that {\it all} $2$-d
models will behave like Ising models in presence of weak disorder. The
purpose of this Letter is to check the validity of this conjecture. More
precisely, if we assume that a model with first order phase transition will
behave like an Ising model in presence of disorder, to how many Ising
models does it correspond ? An easy way to answer this question,
suggested by Cardy, is to measure the central charge of the model in
presence of disorder \cite{cardy}. Then, if this model behaves like $M$
decoupled Ising models, the central charge will simply be $C=M\times
{1\over2}$.

In this Letter we present a study in which we measured the central charge
for some large $q$-state Potts model, in presence of weak disorder
($q=5,8,10,12$ and $256$). It {\it does not} have the central charge of an
integer number of Ising models in general. This is the first result. The
second result is that for $q=2^N$, the central charge is $c_{q=2^N}= N
\times {1\over 2}$. This, in particular, includes the case studied in
\cite{cfl} ($q=8$) for which an Ising-like behavior was obtained. Our
third result is a factorization law. For a general $q$, we
found that if $q=q_1 \times q_2$, then $c_q = c_{q_1}+c_{q_2}$.

Simulations were performed by computing the free energy on long strips with
varying width $L$. It is well known how to relate the free energy to the
central charge in the periodic case \cite{bcn,affleck}
\beq
\label{fc}
f_L = f_\infty + {\pi c \over 6 L^2} + \cdots
\eeq
The free energy being negative, we denote here by $f_L$ the free energy
times $(-1)$, $\ie$
\beq
f_L ={1\over L}\lim_{N\rightarrow \infty}{1\over N}\log{Z_{L,N}}\; .
\eeq
$N$ is the length of the strip and $Z_{L,N}$ the partition function. Thus
this gives us a very fast way of computing the central charge, the only
problem being to compute the free energy on such strips. This can be done
by iterating the transfer matrix. 

We used two different algorithms to perform our measurements. First an
algorithm which iterates the transfer matrix, using a sparse-matrix
factorization with periodic boundary conditions. The advantage of using
sparse-matrix factorization is that it reduces the number of elementary
operations from $q^{2L}$ to $Lq^L$ per iterated row
\cite{nightingale}. This algorithm is quite efficient for small values of
$q$ but for larger values we used another algorithm.  In this second
algorithm, we used the fact that not all the elements of the transfer
matrix are different. Then, starting with equally probable distributed
states, we can only keep the values which are different in our
iteration. The only problem is to construct the different elements of the
transfer matrix. This becomes particularly complicated because of the
presence of the disorder. For instance, for $L=6$, the number of different
elements would be $203$. (The number $N_p(L)$ of such elements for a fixed
width $L$ is given by
\beq
N_p(L)=\sum_{i_2=1}^2 \sum_{i_3=1}^{m_3} \sum_{i_4=1}^{m_4}
\cdots \sum_{i_L=1}^{m_{L-1}} 1
\eeq
with $m_i=max(i_2,i_3,\cdots,m_{i-1})+1$.)  Then we have to compute the
recursion relations between these $203$ elements. This leads us to compute
$203 \times 203 \times 64 $ polynomials (the last $64$ are for the number
of possible configurations of disorder). The advantage of this algorithm
compared to the sparse-matrix one is that it does not depend on the values
of $q$. Then, we can simulate with the same cost in time the free energy
for any $q$-state Potts model.  There is still a limit due to the fast
increase of the number of polynomials. For $L=7$, there would be $877
\times 877 \times 128$ such polynomials. Thus, with this algorithm, we were
only able to perform simulations with a lot of statistics up to $L=6$.

The disorder is simulated by allowing the bonds $J$ to take two values,
$J_0$ and $J_1$, with equal probability. The critical temperature is
determined by solving the equation \cite{kd}
\beq
{1-e^{-\beta J_0} \over 1+(q-1)e^{-\beta J_0}} = e^{-\beta J_1} \; .
\eeq
The strength of the disorder which was chosen is such that
$J_0/J_1=10$. The reasons for such a choice are as follows. It is well
known that if we add disorder to a model with a second order phase
transition, then the strength of the disorder will be related to a
crossover length. This crossover length can be seen like an average
distance between impurities. The stronger the disorder, the smaller the
distance between these impurities. In order to detect the effect of the 
disorder, this average distance should be smaller than the lattice size
that we used. In the case of simulations that we report here, we want this
crossover length to be as close as possible to the lattice spacing. A
direct estimate of this crossover length is too difficult to perform.  So
we have first made simulations for only one value of $q$ ($q=8$) for
different values of the disorder $J_0/J_1$ and measured $f(2)-f(3)$. For a
small disorder (up to $J_0/J_1 \simeq 5$), this value does not change much
from the pure case. Then, we notice a brutal change between $5< J_0/J_1 <
20$.  A similar result was also obtained for $q=5$. A second constraint is
given by the importance of the fluctuations. The stronger the disorder,
the stronger the fluctuations of the measured values of the free
energy will be. The compromise that we made was to take the value $J_0/J_1=10$.

Our simulations have been performed for the following values of $q$ : $q=5,
8, 10, 12$ and $256$. For each of these value of $q$, we have computed the
free energy for $L=2, 3, 4, 5$ and $6$. For $L=2,3$ and $4$ the transfer
matrix was iterated $10^9$ times, for $L=5$ we made $4.10^8$ iterations and
for $L=6$, $2.10^7$ iterations. Errors were computed by separating the date
in $\simeq 1000$ independent runs (for $L=2,3,4$ we made $1000$ runs of
$10^6$ iterations, \etc) and then we took the average over these
independent runs and the error bars were computed by root-mean-square
average of deviations. The values of the free energies that we measured are
reported in Table 1 with the error in the last quoted digits in
parentheses. Some simulations were also performed on larger strip widths
($L=7$ and $8$) for $q=5$ but only over $\simeq 10^6$ iterations. Because
we were not able to accumulate enough data for these larger strip widths, we
will not report about them here, but let us just mention that they give
consistent results. We also performed simulations for other values of $q$
($q=6,11,16,24,64$) but with low statistics. Again, the results for these
values of $q$ will be used to check the consistency of our results.

The next step is to compute the central charge. This was done by performing
a fit with eq. (\ref{fc}).  Then $c$ is obtained by a two-point fit and is
given by
\beq
c_1(q,L)={6\over \pi} {L^2 (L+1)^2 \over 2L+1} (f_L(q) - f_{L+1}(q))\; .
\eeq
The values that we obtain with such a fit are reported in Table 2. We see
that the values of the central charges $c_1$ change with
increasing $L$, as can be expected with such small $L$ and thus we cannot
obtain reliable values for the central charge without increasing the strips
width. At this
point, it is important to notice that  
\beq
\Delta c_1(q,L) \simeq {3\over \pi} L^3 (\Delta f_L(q) + \Delta f_{L+1}(q)) 
\eeq
for large $L$. Thus, for some fixed $\Delta c_1(q,L)/c_1(q,L)$, we need to
measure $f_L(q)$ such that $\Delta f_L(q)\simeq L^{-3}$. But, as explained
above, the time to iterate the transfer matrix increases very quickly with
$L$ and the free energies measured are only very weakly
self-averaging. Thus the best strategy is certainly not to obtain $c(q)$
as the limit of $c_1(q,L)$ for large $L$. A better strategy would be to add
corrections to eq. (\ref{fc}). Bl\"ote and Nightingale, in an extensive
study of the $q$-state Potts models, showed that the fit is improved with a
correction of the following form \cite{bn}
\beq
\label{fc2}
f_L = f_\infty + {\pi c \over 6 L^2} + {b\over L^4} + \cdots
\eeq
and this only for $q\leq 4$, \ie for second order phase transition. (In
\cite{bn}, $a={c\pi \over 6}$ was some constant not yet associated with the
central charge.) For $q>4$, such a fit is completely inconsistent.  In
Table 3, we report the results obtained from a three-point fit of $f(q,L)$
with eq. (\ref{fc2}). To show the relevance of such a fit for small $L$, we
also report the values obtained from the free energy computed by Bl\"ote
and Nightingale for $q=2,3$ and $4$.  First, let us look at the results for
$q=2,3$ and $4$ (without disorder). For these $q$'s, we see that the
deviation of $c_2$ from the real values ($c(2)=0.5$, $c(3)=0.8$ and
$c(4)=1$) is of order $\Delta c/ c \simeq 1/100$ already for a three-point
fit with $L=3-5$ or $L=4-6$.  The original motivation of our study was to
check if $c(q)$ could take half-integer values. For this purpose, a
precision of $1/100$ is more than enough. Now looking at the results for
$c_2(q,L)$ obtained in a three-point fit with $L=4-6$ for $q>4$, we see
that the errors due to the fluctuations are already larger than what is
expected from an extrapolation of the results for $q=2,3$ and $4$. Thus the
values obtained by this three-point fit should be very close of the real
result. By averaging the results from $L=3-5$ and $L=4-6$, we obtain
$c_2(5) = 1.15-1.2, c_2(8)= 1.45-1.55, c_2(10) = 1.6-1.7, c_2(12) =
1.8-1.85$ and $c_2(256) = 4-4.1$

We also have to mention that other type of corrections to
eq. (\ref{fc}) should be incorporated. It has been shown by Cardy that
$\log(L)$ corrections exist, due to the presence of irrelevant
operators \cite{cardy2}. However, these corrections are very small and
should not be important at our level of precision.

Let us now discuss our results. First, for $q=8$, we know that under the
influence of the disorder, the model will behave like an Ising model. We
obtain $c \simeq 1.5$ which means that the $q=8$ state Potts model, in
presence of disorder, behaves like three decoupled Ising models. More
generally, a $q=2^N$ state Potts model will have a central charge
$c(q)=N.{1\over 2}=N c(2)$. We have checked this result by computing the
central charge for $q=256$, $c(256)\simeq 4 = 8 c(2)$ but also for some
over values ($q=16, 64$) for which we have some small data that we will not
report here but which is compatible with this rule (we measured
$c(16)\simeq 2\pm 0.1$ and $c(64)\simeq 3 \pm 0.1$). In addition we
have $c(4)=2 c(2)$, thus the result $c(q=2^N)=N c(2)$ is valid for any
$N$. Then, for $q=5$, we obtained a value which is very far from any
half-integer, $c(5)=1.15-1.2$. This would presumably correspond to a new
conformal field theory. The next result is for $q=10$. For $q=10$, we
obtained $c(10)=1.6-1.7$. Thus we found $c(10)=c(5)+c(2)$ (up to some
large errors). Again, we do not know of any conformal field theory with such
a central charge. (However, a similar value for $c$ has been obtained for
the frustrated $XY$ model ($c=1.66 \pm 0.04$) \cite{tk}.) Thus we observe
the factorization law
\beq
c(q=q_1\times q_2) = c(q_1) + c(q_2) \; .
\eeq
In order to check this factorization law, we have also computed the central
charge for $q=12$. There we obtained $c(12) \simeq 1.8 = 2 c(2) +c(3)$
which corroborate our factorization law. This was also checked for
some other values of $q$ but with smaller data. All the other cases that we
simulated give completely compatible results with this factorization law
(we obtained $c(6) \simeq 1.3 \pm 0.1 \simeq c(2) + c(3)$ and $c(24)\simeq
2.3 \pm 0.1 \simeq 3c(2) + c(3)$).

Then, a further consistency test is to perform a simulation with a
different boundary condition on the strip. If we choose free boundary
conditions, then eq. (\ref{fc}) is replaced by
\beq
f_L = f_\infty + {f_0 \over L} + {c\pi \over 24 L^2} + \cdots
\eeq
We have an additional term (${f_0\over L}$) which corresponds to the
surface free energy. The computation of $c(q)$ is a lot more complicated
because we have first to determine $f_0$ and then to compute the central
charge. Thus the error bar for the central charge would be a lot larger
than for the periodic boundary case. We have performed simulations for
$q=8$ and we have obtained $c(8) \simeq 1-2$, which is compatible with the
result for the same model with periodic boundary conditions.

In conclusion, we have studied the influence of disorder on some $q$-state
Potts models, with $q=5,8,10,12$ and $256$ by measuring the central
charge. We have found that the central charge is not a multiple of ${1\over
2}$ and thus these models {\it do not} behave like multiple Ising models
for a general $q$. But when $q$ is a power of $2$, we do find a
half-integer value for the central charge. For other values of $q$, we
found some new conformal field theories. In addition we found a
factorization law such that $c_{q_1 q_2}=c_{q_1} + c_{q_2}$. All these
measurements were performed on very small strip width, and thus the
precision on which the latter prediction is made is small. More extensive
simulations, and for other values of $q$ should be performed in order to
reinforce this result.

This research was supported in part by National Science Foundation under
Grant No. PHY94-07194. The author thanks the Institute for Theoretical
Physics at Santa Barbara, where this work was started, for its hospitality.


\newpage
\begin{table}
\begin{center}
\begin{tabular}{|l||l|l|l|l|l|} \hline
 L & q=5 & 8 & 10 & 12 & 256   \\
\hline
2 & 3.383531 (31) & 3.825537 (33) & 4.039727 (34) & 4.216639 (34) & 7.367390 (53) \\
\hline
3 & 3.283052 (24) & 3.694779 (26) & 3.894575 (27) & 4.059728 (27) & 7.015583 (40) \\
\hline
4 & 3.249701 (21) & 3.651447 (22) & 3.846513 (23) & 4.007812 (24) & 6.900199 (34) \\
\hline
5 & 3.234857 (30) & 3.632161 (36) & 3.825131 (33) & 3.984714 (33) & 6.849156 (39) \\
\hline
6 & 3.227029 (142) & 3.622131 (130) & 3.813995 (137) & 3.972503 (135) & 6.821969  (199) \\
\hline
\end{tabular}
\end{center}
\protect\caption{\label{table1}Free energy as function of $q$ and $L$.}
\end{table}
\begin{table}
\begin{center}
\begin{tabular}{|l||l|l|l|l|l|} \hline
L & q=5 & 8 & 10 & 12 & 256   \\
\hline
2-3 & 1.3817 (4) & 1.7981 (5) & 1.9960 (5) & 2.1577 (5) & 4.8377 (8) \\
\hline
3-4 & 1.3095 (9) & 1.703 (1) & 1.888 (1) & 2.040 (1) & 4.533 (2) \\
\hline
4-5 & 1.260 (3) & 1.637 (3) & 1.813 (3) & 1.961 (3) & 4.333 (4) \\
\hline
5-6 & 1.23 (2) & 1.57 (2) & 1.74 (3) & 1.91 (3) & 4.25 (3) \\
\hline
\end{tabular}
\end{center}
\protect\caption{\label{table2} Central charge $c_1(q)$ from a fit with
Eq. (\ref{fc}).} 
\end{table}
\begin{table}
\begin{center}
\begin{tabular}{|l||l|l|l||l|l|l|l|l|} \hline
L & q=2 & 3 & 4 & 5 & 8 & 10 & 12 & 256   \\
\hline
2-4 & 0.53004 & 0.85335 & 1.08010 & 1.243 (3) & 1.614 (3) & 1.789 (3) & 1.930 (3) & 4.251 (4) \\
\hline
3-5 & 0.49875 & 0.79846 & 1.00493 & 1.189 (8) & 1.543 (8) & 1.709 (8) & 1.847 (8) & 4.044 (9) \\
\hline
4-6 & 0.49507 & 0.79222 & 0.99610 & 1.16 (5) & 1.45 (6) & 1.60 (7) & 1.81 (7) &  4.08 (9) \\
\hline
\end{tabular}
\end{center}
\protect\caption{\label{table3} Central charge $c_2(q)$ from a fit with
Eq. (\ref{fc2}).} 
\end{table}
\end{document}